\documentclass[twocolumn]{emulateapj}

\usepackage{graphicx}
\usepackage{epsfig}
\usepackage{amsmath}

\usepackage{color}

\shorttitle{Rotational Corrections to Neutron-Star Spectra}
\shortauthors{Baub\"ock et al.}

\begin{document}


\title{Rotational Corrections to Neutron-Star Radius Measurements from Thermal Spectra}

\author{Michi Baub\"ock\altaffilmark{1}, Feryal \"Ozel\altaffilmark{1}, Dimitrios Psaltis\altaffilmark{1}, and Sharon M.\ Morsink\altaffilmark{1,2}}
\altaffiltext{1}{Astronomy Department,
University of Arizona,
933 N.\ Cherry Ave.,
Tucson, AZ 85721, USA}
\altaffiltext{2}{Permanent address:
Department of Physics,
University of Alberta,
11455 Saskatchewan Drive,
Edmonton, AB T6G 2E1, Canada}
\email{mbaubock@email.arizona.edu}

\begin{abstract}
We calculate the rotational broadening in the observed thermal spectra of neutron stars spinning at moderate rates in the Hartle-Thorne approximation. These calculations accurately account for the effects of the second-order Doppler boosts as well as for the oblate shapes and the quadrupole moments of the neutron stars. We find that fitting the spectra and inferring the bolometric fluxes under the assumption that a star is not rotating causes an underestimate of the inferred fluxes and, thus, radii. The correction depends on the stellar spin, radius, and observer's inclination. For a 10~km neutron star spinning at 600~Hz, the rotational correction to the flux is $\sim 1-4\%$, while for a 15~km neutron star with the same spin period, the correction ranges from 2\% for pole-on sources to 12\% for edge-on sources. We calculate the inclination-averaged corrections to inferred radii as a function of the neutron-star radius and mass and provide an empirical formula for the corrections. For realistic neutron star parameters (1.4~M$_\odot$, 12~km, 600~Hz), the stellar radius is on the order of 4\% larger than the radius inferred under the assumption that the star is not spinning.
\end{abstract}


\section{Introduction}

Neutron stars spin at high frequencies, as revealed by pulsar emission in the radio wavelengths as well as pulsed emission and burst oscillations in the X-rays. The distribution of spin frequencies measured through these phenomena currently extends up to 716~Hz (Hessels et al.\ 2006). Such high spin frequencies correspond to surface velocities that exceed a tenth of the speed of light, which give rise to large Doppler effects. As these velocities become comparable to the stellar break-up velocities, the rapid spins also distort the shape of the neutron-star surfaces and modify a broad range of observable characteristics. 

The spectrum of emission from the stellar surface is one of the observables affected by rapid stellar spins. To first order in spin frequency, Doppler shifts experienced by photons emitted from the surface broaden the spectrum while preserving the total flux. To second order, the stellar surface becomes oblate and the star acquires a non-zero quadrupole moment (Hartle \& Thorne 1968). Both of these effects cause different parts of the stellar surface to experience different gravitational redshifts, further distorting both line and continuum spectra.

Baub\"ock et al.\ (2013a) calculated the effects of moderate stellar spin on the profiles of narrow atomic features. They found that for certain regions of the parameter space, the second-order effects of oblateness and the associated quadrupole moment significantly affect line profiles (see also Psaltis \& \"Ozel 2014 for similar effects in the context of pulse profiles).  In particular, moderately spinning neutron stars observed at low inclinations ($\le 30^\circ$) display unexpectedly narrow and asymmetric line profiles.

Observations of rotationally broadened line profiles lead, in principle, to measurements of neutron star masses and radii (\"Ozel \& Psaltis 2003; Chang et al.\ 2006; Bhattacharyya et al.\ 2006). However, no line features originating from a neutron star surface have been definitively detected (see Cottam et al.\ 2002, 2008; Lin et al.\ 2010). On the other hand, many classes of sources display broadband thermal emission and offer one of the best methods for measuring the radii of neutron stars. Since quantities such as the pressure and density in stellar cores are not directly observable, these radius measurements, in turn, provide one of the best ways to constrain the equation of state of cold dense matter (see, e.g., Lattimer \& Prakash 2000; Read et al.\ 2009; \"Ozel \& Psaltis 2009; \"Ozel 2013 and references therein).

Thermal emission has been used to constrain neutron-star masses and radii in the context of thermonuclear X-ray bursts since the discovery of bursters in the 1970s (e.g., van Paradijs 1979). For pure blackbody emission from a uniform spherical star, the spectroscopic radius of of a star can be calculated from the Stefan-Boltzmann law if the star's flux, temperature, and distance are measured. More recently, this method has been applied to bursting sources using high precision Rossi X-ray Timing Explorer data (e.g., \"Ozel et al.\ 2009, 2010; G\"uver et al.\ 2010). In addition, thermal emission from transient low-mass X-ray binaries during quiescence (e.g., Heinke et al.\ 2006, 2014; Webb \& Barret 2007; Guillot et al.\ 2013; Catuneau et al.\ 2013) as well as from isolated neutron stars (e.g., Pons et al 2002; Drake et al.\ 2002) has been observed and used for spectroscopic determination of neutron star radii.

Measurements of neutron-star radii need to reach an accuracy of 5-10\% in order to distinguish between different proposed equations of state and thus help determine the physical processes that play a role in neutron-star interiors (Lattimer \& Prakash 2000; \"Ozel \& Psaltis 2009). A number of measurement and modeling uncertainties need to be pinned down and overcome in order for this required level of accuracy to be reached.  Earlier work has focused on several of these effects, which we summarize below. 

On the measurement side, G\"uver et al.\ (2012) showed that the statistical error as well as the systematic uncertainties in the apparent angular sizes arising from possible non-uniform emission across the surface of neutron stars are at or below 10 percent for X-ray burst sources. The systematic uncertainties of the absolute flux calibration of X-ray detectors were similarly shown to be below the 10 percent level (G\"uver et al. 2014 in prep).

An important component of theoretical modeling is related to spectral distortions caused by radiative transfer effects in the neutron-star atmosphere. Because the free-free opacity depends strongly on photon energy, the color temperature measured from the peak of the spectrum emerging from the atmosphere is expected to be significantly higher than the effective temperature of the surface (van Paradijs 1979). In addition, the Compton scattering of photons off of the free electrons in the atmosphere further distort the emitted spectra (London et al.\ 1986). Accurate models of neutron-star atmospheres have accounted for these effects for a wide range of stellar parameters (London et al.\ 1986; Madej et al.\ 2004; Suleimanov et al.\ 2012) and the results of these calculations have been converging and now agree to within 10 percent (see the discussion in \"Ozel 2013).

In this paper, we investigate the effects of moderate to fast rotation on thermal spectra emitted from neutron-star surfaces and address this remaining source of uncertainty in radius measurements. It is well known that gravitational redshifts and strong-field gravitational lensing along the photon paths between the stellar surface and the observer play a significant role in determining the angular size of the source measured spectroscopically. For slowly spinning stars, these effects can be calculated analytically in the case of uniform emission from the stellar surface (Pechenick et al.\ 1983). In this limit, the apparent spectroscopic radius is larger than the circumferential radius by a factor of $1/\sqrt{1 - 2 G M / R c^2}$, where $M$ and $R$ are the mass and radius of the neutron star. All previous determinations of neutron star radii using the spectroscopic method have utilized this approximation. In this paper, we examine the effect of rotation on measurements of the spectroscopic radius.

Two previous studies have examined the effects of rotational broadening on thermal spectra emitted from the surfaces of neutron stars. Asaoka \& Hoshi (1987) and Fu \& Taam (1990) calculated the Doppler broadening of thermal spectra using the Kerr and Schwarzschild metrics, respectively. Both studies concluded that rotational effects were unimportant for stars spinning below $\sim 500$~Hz, which encompassed the range of spin frequencies observed at the time. Thus, they did not consider corrections to the spectroscopic radius. However, using the Hartle-Thorne metric, Baub\"ock et al.\ (2012) showed that even moderate stellar spins affect the propagation of photons from the stellar surface to an observer at infinity, distorting the apparent shape and size of the stars.

We calculate simulated spectra from moderately spinning sources using an algorithm (Baub\"ock et al.\ 2012, 2013a) that traces rays in a variant of the Hartle-Thorne metric (Hartle \& Thorne 1968). We investigate the distortions in the broadband spectra resulting from spin effects, taking into account the oblate shape of the stellar surface and the quadrupole moment of the metric. Using these model spectra, we calculate correction terms to account for the effects of rotation on measurements of the effective temperature, flux, and spectroscopic radius that are accurate for spin frequencies up to 800~Hz. These can be used to infer neutron-star radii from moderately spinning sources.

\section{Spectral Modeling}

We use the ray-tracing algorithm described in Baub\"ock et al.\ (2012, 2013a) to calculate spectra observed at infinity for a range of neutron-star parameters. We model neutron stars in a variant of the Hartle-Thorne metric (Hartle \& Thorne 1968; see Glampedakis \& Babak 2004), which allows for neutron stars with a moderate spin frequency to assume an oblate shape and the exterior spacetime to have an appropriate quadrupole moment.

In principle, the metric of the spacetime around the neutron star depends on six parameters: the mass $M$, the equatorial radius $R$, the spin frequency $f_{\rm NS}$, the moment of inertia $I$, the quadrupole moment $Q$, and the ellipticity of the surface of the star $\varepsilon$. Additionally, measurements of the source depend on the inclination of the spin axis to the observer's line of sight $\theta_O$. Using empirical and analytic relations (Baub\"ock et al.\ 2013b; Morsink et al.\ 2007; see also AlGendy \& Morsink 2014), we determine three of these parameters ($I$, $Q$, and $\varepsilon$) in terms of only the mass, equatorial radius, and spin frequency.

We model neutron stars with masses between 1.0 and 2.5~M$_\odot$, radii between 8 and 16~km, and spin frequencies between 50 and 800~Hz over all possible inclinations. For each configuration, we produce a transfer function between the spectrum at the surface of the star and the spectrum an observer measures at a distant location, which we define below.

Photons are initialized on an image plane located in the asymptotically flat spacetime at a large radial distance from the neutron star. The photons are then traced backwards until they intersect the stellar surface. We record the position and energy of each photon on the surface and use these to find an image of the source on the image plane. We then integrate the specific intensity over this image in order to find the flux measured by an observer at spatial infinity as
\begin{equation}
F_{\infty}(E_{\infty}) = \frac{1}{D^2} \iint \! I_{\infty}(E_{\infty}, \alpha, \beta) \, 
\mathrm{d} \alpha \, \mathrm{d} \beta,
\label{eq:A_Flux_1}
\end{equation}
where $E_\infty$ is the energy measured at infinity, $D$ is the distance to the neutron star, $I_\infty$ is the specific intensity measured at infinity, and $\alpha$ and $\beta$ are Cartesian coordinates on the image plane (see Baub\"ock et al.\ 2012 for detailed definitions).

Since the quantity $I/E^3$ is Lorentz invariant, we can write
\begin{equation}
\frac{I_\infty(E_\infty)}{E_\infty^3} = \frac{I_s(E_s)}{E_s^3},
\label{eq:A_Lorentz}
\end{equation}
where $I_s$ and $E_s$ are the specific intensity and energy, respectively, as measured on the surface of the neutron star. We assume uniform and isotropic emission over the surface of the star, i.e., that $I_s$ does not depend on $\alpha$ and $\beta$. If we define 
\begin{equation}
g(\alpha, \beta) \equiv E_\infty/E_s,
\label{eq:g_def}
\end{equation}
we can then write equation (\ref{eq:A_Flux_1}) as
\begin{equation}
F_\infty(E_\infty) = \frac{1}{D^2} \iint \! g^3 I_s\left(\frac{E_\infty}{g} 
\right) \, \mathrm{d} \alpha \, \mathrm{d} \beta.
\label{eq:A_Flux_2}
\end{equation}

Our goal is to find a transfer function $G(E_\infty, E_s)$ such that we can write the flux at infinity as
\begin{equation}
F_\infty(E_\infty) = \int \! I_s(E_s) G(E_\infty, E_s) \, \mathrm{d} E_s.
\label{eq:A_Flux_3}
\end{equation}
This allows us to understand the effects of rapid spin independent of the source spectrum. In order to convert equation~(\ref{eq:A_Flux_2}) into this form, we need to translate the integral over the image plane into an integral over the energies on the stellar surface. We do this by introducing a Dirac delta function which is zero at all points on the image plane except at those where the energy on the surface $E_s$ is equal to the energy at infinity $E_\infty$ scaled by the factor $g$, $\delta(E_\infty/g - E_s)$. Since delta functions have the property that 
\begin{equation}
\delta(a x) = \frac{1}{|a|} \delta(x), 
\label{eq:del_prop}
\end{equation}
we can rewrite this as
\begin{equation}
\delta\left(\frac{E_\infty}{g} - E_s\right) = g \delta(E_\infty - g E_s). 
\label{eq:del_g}
\end{equation}
and equation~(\ref{eq:A_Flux_2}) as
\begin{equation}
F_\infty(E_\infty) = \frac{1}{D^2} \iiint \! I_s(E_s) g^4 \delta(E_\infty - g E_s) \, 
\mathrm{d} E_s \, \mathrm{d} \alpha \, \mathrm{d} \beta.
\label{eq:A_Flux_4}
\end{equation}
By comparing equation~(\ref{eq:A_Flux_3}) to equation~(\ref{eq:A_Flux_4}), we find the transfer function
\begin{equation}
G(E_\infty, E_s) = \frac{1}{D^2} \iint g^4 \delta(E_\infty - g E_s) 
\, \mathrm{d} \alpha \, \mathrm{d} \beta.
\label{eq:G}
\end{equation}

To calculate an observed spectrum for a given neutron-star configuration, we first calculate the transfer function $G(E_\infty, E_s)$ from equation~(\ref{eq:G}). We then use this function as a kernel for a convolution with the spectrum at the surface, as in equation~(\ref{eq:A_Flux_3}). In principle, the surface spectrum $I_s(E_s)$ can consist of a detailed atmospheric model, but for the purposes of this paper we restrict our analysis to a blackbody function with an effective temperature $T_s$. 

\begin{figure}
\psfig{file=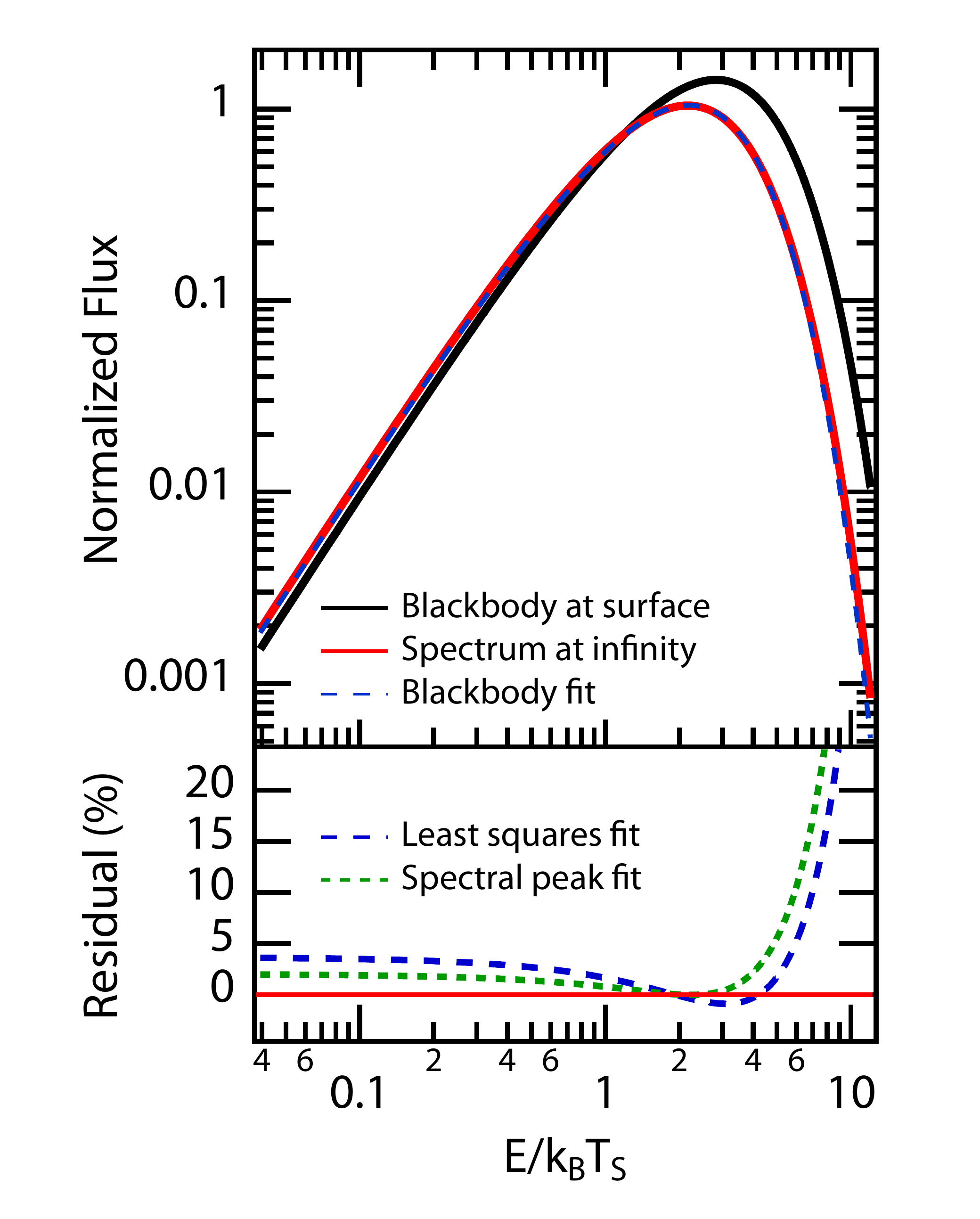, width=3.5in}
\caption{Broadened blackbody spectrum of temperature $T_s$ emitted from the surface of a neutron star of mass 1.4~$M_{\odot}$ and radius 10~km, spinning at 700~Hz, as observed at an inclination of 85 degrees. The x-axis shows photon energy scaled by $k_{\rm B} T_s$. The surface spectrum has been normalized by a constant factor such that it has an integral of 1. The black solid line shows the blackbody spectrum of temperature $T_s$ measured at the surface of the star. The red line shows the broadened spectrum measured at spatial infinity. The blue dashed line shows the best fit blackbody to the broadened spectrum found by a least-squares fitting algorithm in the energy range $0.04 \le E/k_{\rm B} T_s \le 12$. The curve corresponding to a different fit that matches the energy and flux of the  spectral peak is indistinguishable from the blue dashed line in the upper panel. The lower panel shows the residuals for both fits. Rotational effects introduce spectral distortions at the $\lesssim 5\%$ level.}
\label{fig:BB_Broadened}
\end{figure}

The top panel of Figure~\ref{fig:BB_Broadened} shows the blackbody spectrum as measured in the rest-frame on the surface of the neutron star as a solid black curve. The spectrum measured by an observer at infinity for a star spinning at 700~Hz is shown as a solid red curve.

As expected, the spectrum at infinity is both broadened and shifted with respect to the surface spectrum and can no longer be represented by a blackbody. However, for the parameters of this calculation, the deviation from a blackbody spectrum is relatively small. In order to find the approximate blackbody temperature that would be measured at infinity, we fit a blackbody to the observed spectrum in two different ways. First, we locate the peak of the broadened spectrum and calculate the temperature and normalization of a blackbody with the same peak energy and flux. Second, we perform a fit to find the parameters of a blackbody with the minimum least-squares difference to the broadened spectrum over the range of energies considered (0.04--12~$k_{\rm B}T_s$). The dashed line in the upper panel in Figure~\ref{fig:BB_Broadened} shows the latter fit, and the lower panel shows the residuals for both fitting methods. As the two procedures produce the same blackbody parameters to within a percent, we use only the least-squares method for the remainder of this work. The deviation of the spectrum at infinity from a blackbody varies from 3--4\% below the peak to $\ge$ 5\% at higher energies. This is comparable to the measured deviation of observed spectra from blackbodies in X-ray bursters (G\"uver et al.\ 2012). 

Both the measured flux and the blackbody temperature are used in inferring the masses and radii of thermally emitting neutron stars. When these inferences are made under the assumption of a non-rotating source, a systematic error is introduced in the measurements. For a non-spinning neutron star, the apparent spectroscopic radius $R_\infty$ can be calculated from the measured flux $F_\infty$ and temperature $T_\infty$ using
\begin{equation}
F_\infty = \sigma \left(\frac{R_\infty}{D}\right)^2 T_{\infty}^4,
\label{eq:f_sch}
\end{equation}
where $D$ is the distance to the source, which we set to unity for the remainder of this work. In this case, the observed spectrum is a blackbody with a redshifted peak and diminished flux, both caused by the gravitational redshift from the stellar surface. In the Schwarzschild metric, both of these corrections can be calculated analytically. The peak of the spectrum (and, therefore, its inferred temperature) is redshifted by
\begin{equation}
\frac{T_\infty^{(f_{\rm NS} = 0)}}{T_s} = \sqrt{1 - \frac{2 G M}{R c^2}}.
\label{eq:T_redshift}
\end{equation}
Likewise, we can find the bolometric flux at infinity by integrating the spectrum and use a similar expression to correct for the gravitational redshift and strong-field lensing, i.e.,
\begin{equation}
\frac{F_\infty^{(f_{\rm NS} = 0)}}{F_s} = \left(1 - \frac{2 G M}{R c^2} \right),
\label{eq:F_redshift}
\end{equation}
where $F_s$ is the flux measured on the surface. This will allow us to relate the spectroscopic radius observed at infinity to the circumferential radius of the star by
\begin{equation}
R = R_\infty \sqrt{ 1 - \frac{2 G M}{R c^2}}.
\label{eq:sch_r}
\end{equation}

For a spinning neutron star, we can compare the values of the inferred temperature and flux to those derived under the assumption that the source is spherical and non-spinning in order to calculate correction factors introduced by the spin. In particular, we define the rotational temperature correction factor as
\begin{equation}
\zeta_{\rm rot} \equiv \frac{T_\infty}{T_\infty^{(f_{\rm NS} = 0)}} = \frac{T_\infty}{T_s \sqrt{1 - \frac{2 G M}{R c^2}}},
\label{eq:f_rot}
\end{equation}
and the correction to the bolometric flux as
\begin{equation}
b_{\rm rot} \equiv \frac{F_\infty}{F_\infty^{(f_{\rm NS} = 0)}} = \frac{F_\infty}{F_s \left(1 - \frac{2 G M}{R c^2}\right)}.
\label{eq:flux_rot}
\end{equation}
These correction factors quantify the bias in calculating the above quantities from an observed spectrum when it is assumed that the source is not rotating. 
Using equations (\ref{eq:f_rot}) and (\ref{eq:flux_rot}) we find that, for a spinning star, the inferred spectroscopic radius will be biased by a factor of
\begin{equation}
\frac{R_\infty}{R_\infty^{f_{\rm NS} = 0}} = \sqrt{\frac{b_{\rm rot}}  {\zeta_{\rm rot}^4} }.
\label{eq:R_correction}
\end{equation}

\section{Results}

\begin{figure}
\psfig{file=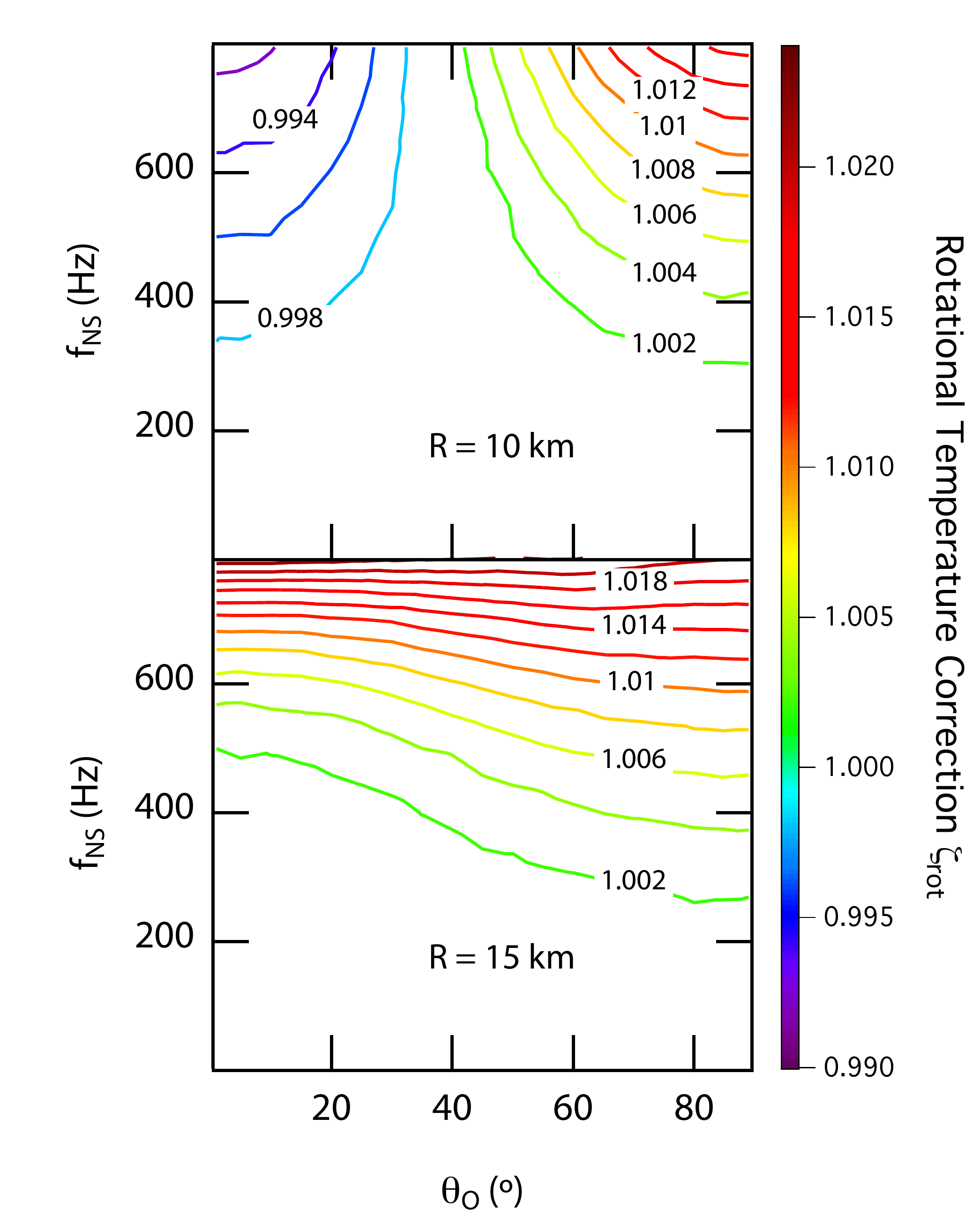, width=3.5in}
\caption{Contours of constant rotational correction to temperature for a range of neutron-star spin frequencies ($f_{\rm NS}$) and inclinations ($\theta_O$). The correction to the temperature is defined as in equation~(\ref{eq:f_rot}). The upper panel shows neutron stars with equatorial radii of 10~km, while the lower panel shows stars with 15~km radii. In both cases, the stars have a mass of 1.4~$M_{\odot}$.}
\label{fig:fc_Contours}
\end{figure}

Figure~\ref{fig:fc_Contours} shows the bias $\zeta_{\rm rot}$ in the measured temperature as a result of the spin of the neutron star. The lines represent contours of constant rotational correction, as defined in equation~(\ref{eq:f_rot}), for a range of possible spin frequencies $f_{\rm NS}$ and inclination angles $\theta_O$. These corrections are small, at the 1\% level, for both values of the neutron star radius shown in this figure. As the spin frequency approaches zero, the measured temperature approaches the value calculated in the Schwarzschild approximation, such that the correction factor becomes unity for all inclination angles.

The shape of the contours and whether the rotational correction factor is smaller or larger than unity for a set of parameters can be understood as follows. At high spin frequencies, Doppler shifts due to the velocity of the stellar surface become significant, and the neutron star becomes oblate in shape. We can determine the effect of Doppler shifts by considering the photon energy $E_\infty$ at infinity from a point on the stellar surface. Ignoring gravitational redshift, this energy is 
\begin{equation}
E_\infty = \eta E_s,
\label{eq:Dopp_shift}
\end{equation}
where $E_s$ is the photon energy at the surface and  $\eta$ is the Doppler function,
\begin{equation}
\eta = \frac{\sqrt{1 - (v/c)^2}}{1 - (v/c) \cos{\xi}}.
\label{eq:Dopp_func}
\end{equation}
In this equation, $v$ is the local magnitude of the velocity of a point on the stellar surface and $\xi$ is the angle between the velocity vector and the direction of emission of the photon. The angle $\xi$ is defined so that $\cos \xi$ is positive on the blueshifted side of the star and negative on the redshifted side. For small values of $v$, the Doppler factor can be approximated by
\begin{equation}
\eta \approx 1 + \frac{v}{c} \cos{\xi} + \left(\frac{v}{c}\right)^2 \left(\cos^2{\xi} - 0.5 \right) + O\left[\left(\frac{v}{c}\right)^3\right].
\label{eq:Dopp_approx}
\end{equation}
When the photon energies are averaged over the surface of the star, the first order term (as well as all higher odd powers of $v/c$) will cancel, leaving only the second order term. At high inclinations, $\cos{\xi}$ is large for most of the stellar surface, leading to a higher energy measured at infinity. At low inclinations, however, $\cos{\xi}$ is smaller than 0.5 for most of the surface, resulting in an average Doppler factor smaller than 1. Moreover, at low inclinations the oblate shape of the surface leads to a stronger gravitational redshift at the pole, further decreasing the inferred temperature.

The changes in the temperature also depend on the compactness of the neutron star. Stars with a smaller compactness (corresponding to a larger radius for a constant mass) have a larger quadrupole moment for a given spin frequency. For a 15~km star, the high quadrupole moment results in a smaller gravitational redshift near the pole, leading to a higher temperature for sources observed at low inclinations, as shown in the lower panel of Figure~\ref{fig:fc_Contours}.

\begin{figure}
\psfig{file=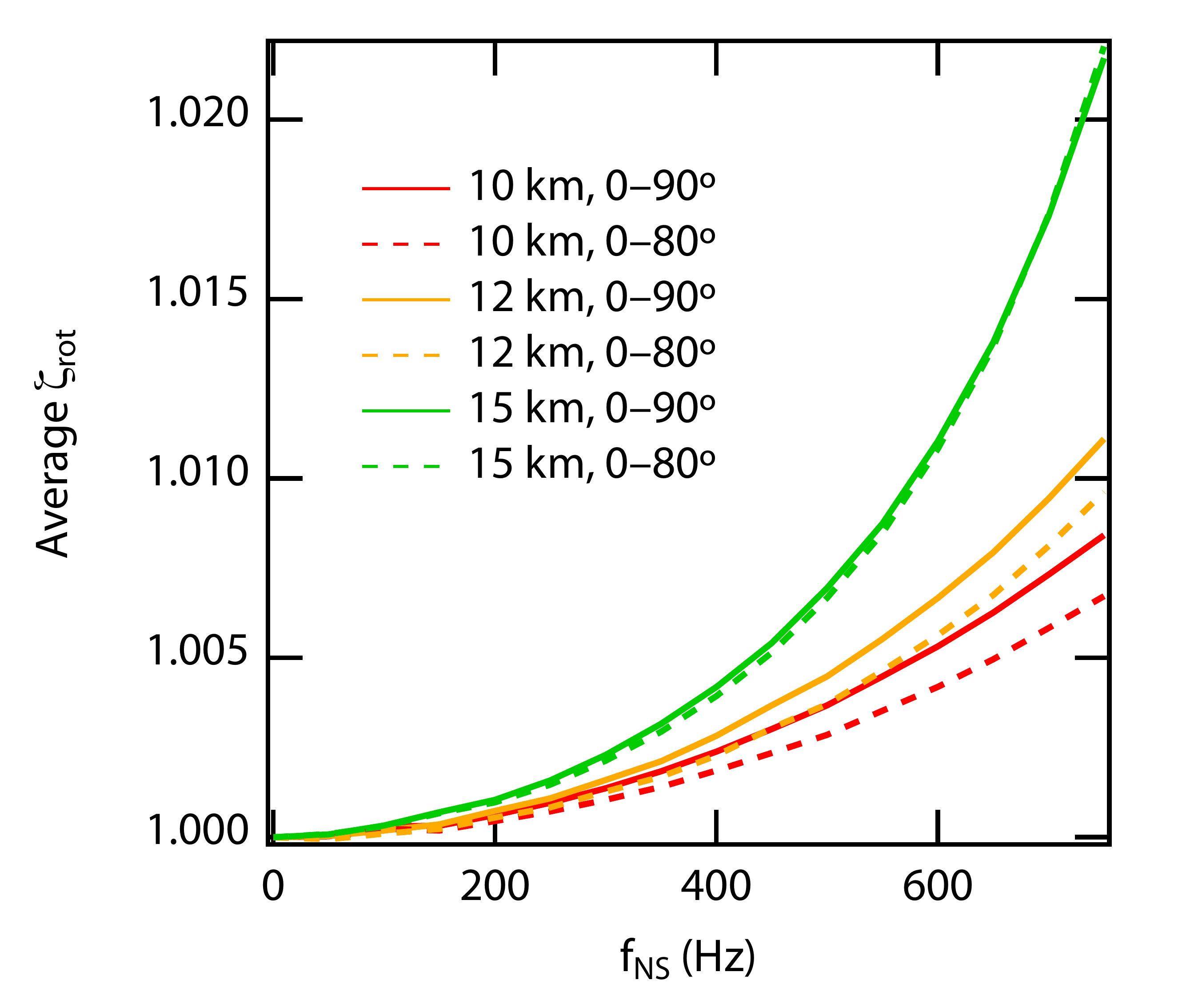, width=3.5in}
\caption{Rotational correction to the neutron-star temperature averaged over inclination. The plot shows the rotational correction to the temperature weighted by the sine of the inclination angle and averaged over a range of possible inclinations for a neutron star with a mass of 1.4~M$_\odot$ and three different radii. The solid lines show the correction averaged over inclinations from $0^\circ$ (pole-on) to $90^\circ$ (equatorial), while the dotted lines show the correction averaged between $0^\circ$ and $80^\circ$.The largest radii lead to the largest corrections.}
\label{fig:fc_sinth}
\end{figure}

For most sources displaying thermal emission, the inclination of the spin axis to the observer's line of sight is not known. In Figure~\ref{fig:fc_sinth} we show the rotational correction to temperature as a function of spin frequency averaged over the inclination of the source assuming a random orientation of the observer. We choose two inclination ranges for these averages: from $0^\circ$ (pole-on) to $90^\circ$ (equatorial) and from $0^\circ$ to $80^\circ$. We calculate the average over the latter range to take into account the possibility that an accretion disk may obscure the surface at high inclinations and, therefore, these objects would not be detected in a search for thermal emission. For both inclination ranges, the correction to the temperature due to the spin of the neutron star is less than 2~\% across the range of observed spin frequencies.

\begin{figure}
\psfig{file=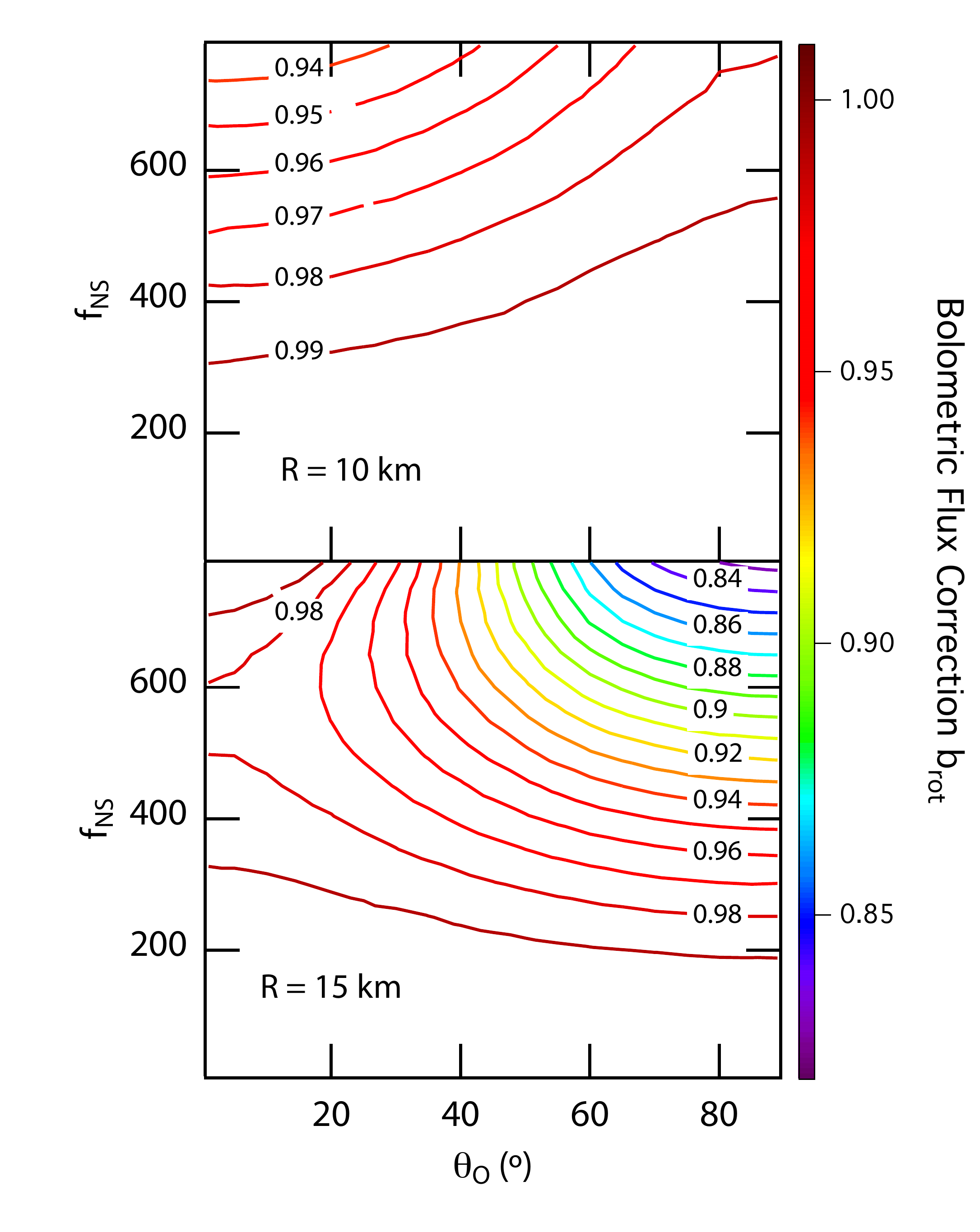, width=3.5in}
\caption{Contours of constant rotational correction to the bolometric flux for a range of neutron star spin frequencies $f_{\rm NS}$ and inclinations $\theta_O$. The correction to the flux is defined as in equation~(\ref{eq:flux_rot}). The upper panel shows the result for a neutron star with a 10~km radius, while the lower panel shows a 15~km star. Both stars have a mass of 1.4~$M_{\odot}$.}
\label{fig:N_Contours}
\end{figure}

In Figure~\ref{fig:N_Contours}, we show the correction to the bolometric flux due to the neutron-star spin as defined in equation~(\ref{eq:flux_rot}). As before, the zeroth order gravitational redshift has been removed, and first order Doppler shift effects cancel when integrated over the surface.  We find that $b_{\rm rot}$ is less than 1 across the parameter space,  indicating that rapid rotation decreases the flux measured by an observer at infinity. The value of $b_{\rm rot}$ depends on several competing effects at second order in $v/c$: the second order terms of the Doppler shift, the oblate shape of the star, the quadrupole moment of the spacetime, and the frame dragging near the surface. We can first gain some insight into the relative magnitudes of these effects by calculating typical values for these corrections and how they scale with the neutron-star parameters.  

\begin{figure*}
\psfig{file=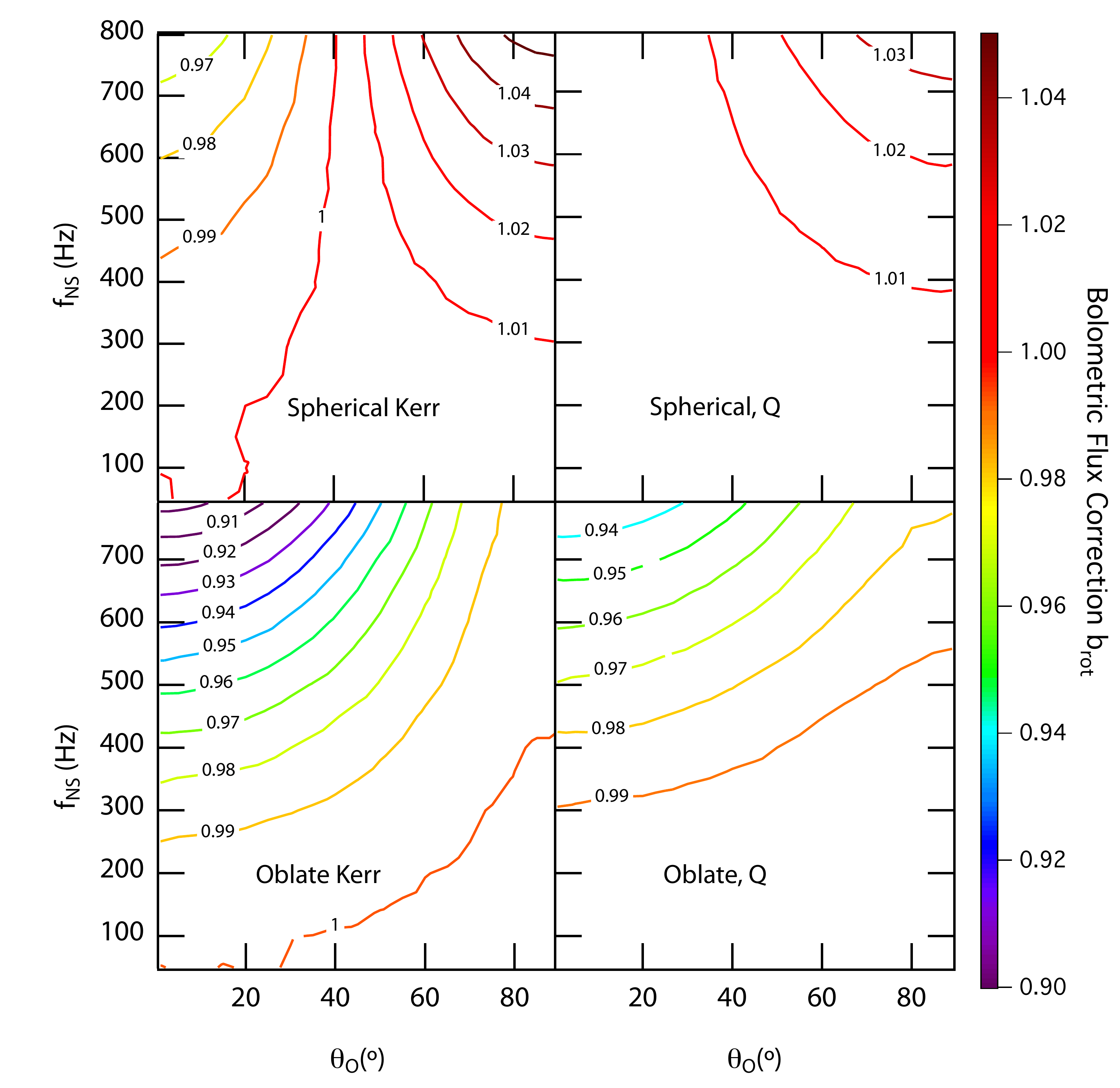, width=6.5in}
\caption{Contours of constant rotational correction to the bolometric flux for four different neutron-star configurations aiming to illustrate the effects of surface geometry and quadrupole moment. Each panel shows the result for neutrons stars with a mass of 1.4~$M_\odot$ and a 10~km equatorial radius. The top left panel shows the result for a spherical neutron star in the Kerr metric. The top right panel shows the results for a spherical star with an appropriate quadrupole moment. The lower left panel corresponds to a star in the Kerr metric with an oblate surface. Finally, the lower right panel corresponds to a star in the Hartle-Thorne approximation, with an oblate surface and an appropriate quadrupole moment. The lower right panel is identical to the upper panel of Figure~\ref{fig:N_Contours}. }
\label{fig:4_comp}
\end{figure*}  

We estimate the contributions of the Doppler shift by integrating equation~(\ref{eq:Dopp_shift}) for a spherical star in Newtonian gravity at an inclination of $90^\circ$,
\begin{equation}
\frac{\Delta E_\infty}{E_\infty} = \int_0^\pi \int_0^\pi \! \eta \, \mathrm{d}\theta \mathrm{d} \phi
\label{eq:zeta_analytic}
\end{equation}
We find that the average Doppler shift is 
\begin{equation}
\frac{\Delta E_\infty^{\rm Doppler}}{E_\infty} \approx 0.004 \left(\frac{R_{\rm eq}}{10 \, {\rm km}}\right)^2 \left(\frac{f_{\rm NS}}{600 \, {\rm Hz}}\right)^2,
\label{eq:scale_Doppler}
\end{equation}   
where $R_{\rm eq}$ is the equatorial radius. 

Both the oblateness and the quadrupole moment affect the flux and temperature primarily by changing the gravitational redshift at different points on the surface. In the case of oblateness, we start from the Schwarzschild expression for the gravitational redshift 
\begin{equation}
\frac{E_\infty}{E_s} = \sqrt{1 - \frac{G M}{R_{\rm eq} c^2}},
\label{eq:grav_red}
\end{equation}
and estimate the difference in the polar and equatorial redshifts as
\begin{equation}
\frac{\Delta E_\infty^{\rm Oblate}}{E_\infty} \approx - \frac{G M}{R_{\rm eq} c^2} \left(1 - \frac{G M}{R_{\rm eq} c^2}\right)^{-1} \left(1 - \frac{R_{\rm p}}{R_{\rm eq}}\right),
\label{eq:dEE_oblate}
\end{equation}
where $R_{\rm p}$ is the radius at the rotational pole. Using equation~(22) of Baub\"ock et al.\ (2013b), we find that  
\begin{equation}
\frac{\Delta E_\infty^{\rm Oblate}}{E_\infty} \approx -  0.015 \left(\frac{R_{\rm eq}}{10 \, {\rm km}}\right)^2 \left(\frac{f_{\rm NS}}{600 \,{\rm Hz}}\right)^2.
\label{eq:scale_oblate}
\end{equation}

Similarly, to estimate the effect of a non-zero quadrupole moment, we find the gravitational redshift at the poles and the equator using
\begin{equation}
\frac{E_\infty}{E_s} = \sqrt{g_{tt}},
\label{eq:grav_redQ}
\end{equation}
where $g_{tt}$ is the $tt$-component of the metric. The exact form of $g_{tt}$ can be found in Glampedakis \& Babak (2006). Approximating the quadrupole moment with equation~(25) of Baub\"ock et al.\ (2013b), we find that
\begin{equation}
\frac{\Delta E_\infty^{Q}}{E_\infty} \approx  0.002 \left(\frac{R_{\rm eq}}{10 \, {\rm km}}\right)^2 \left(\frac{f_{\rm NS}}{600 \,{\rm Hz}} \right)^2 .
\label{eq:scale_q}
\end{equation}

Each of these effects introduce a bias in the flux and temperature measurements at second order in the spin frequency of the star. For most of the parameter space, the dominant effect is the oblateness, which tends to decrease the observed flux of the star. However, the actual contribution of each bias depends on the inclination of the observer and the gravitational lensing, which depends on $M/R$. Therefore, it is possible for different effects to dominate depending on the neutron-star mass, radius, and inclination. It should be noted that, although the effects calculated above are at or less than the 1\% level, the bias in the flux depends on these factors to the fourth power and can therefore be considerably larger.  

In order to illustrate these effects, we calculated fluxes for several neutron-star configurations that emphasize different effects at second order. (Note that this is only for the purposes of delineating different effects; in reality, an oblate star would always have a non-zero quadrupole moment, etc.) Each of the four panels of Figure~\ref{fig:4_comp} shows the flux correction for a different neutron star configuration and for a range of spin frequencies and observer inclinations. The upper left panel shows the flux for a spherical neutron star in the Kerr metric. In this case, the only effects causing a change in the flux are frame dragging and higher order terms in the Doppler shift. We find that the frame dragging has a negligible effect on the flux. At low inclinations, the transverse Doppler shift causes the limbs of the neutron star to appear redder and, therefore, dimmer, causing the flux correction to be less than 1. At high inclinations, on the other hand, Doppler boosting causes the blueshifted side of the star to appear brighter than the redshifted side. Since the Doppler boost is second order in $v/c$, the decreased flux from the redshifted side does not cancel the increased flux from the blueshifted side of the star, leading to a flux correction greater than 1.

In the bottom left panel, we allow the surface of the neutron star to become oblate with increasing spin frequency while retaining the Kerr metric. Comparing this panel to the Kerr panel, we find that introducing oblateness decreases the measured flux at all spin frequencies and inclinations. At low inclinations, this is due to the additional redshift at the pole, which is deeper in the gravitational well (see equation~[\ref{eq:scale_oblate}]). At high inclinations, the geometric size of the neutron star decreases as it becomes more oblate, also decreasing the measured flux (see Baub\"ock et al.\ 2013a).

The top right panel of Figure~\ref{fig:4_comp} shows a spherical star with a spacetime with a quadrupole moment that is appropriate for a spinning neutron star. Note that this is not a self-consistent calculation since the proper radius of the pole and the equator are not the same. However, we use it here only to isolate the effect of the spacetime quadrupole. Compared to the spherical star in the upper left panel, the quadrupole moment decreases the gravitational redshift at the poles while increasing it at the equator (see equation~[\ref{eq:scale_q}]). Consequently, the flux increases for neutron stars observed at low inclinations compared to the spherical Kerr case and decreases at high inclinations.

Finally, the lower right panel shows the combination of all the above mentioned effects. This panel shows an oblate star in the Hartle-Thorne metric and is identical to the upper panel of Figure~\ref{fig:N_Contours}. At low inclinations, the flux-decreasing effects of the oblate shape and transverse Doppler shift outweigh the decrease in redshift from the quadrupole moment, leading to a flux correction factor less than 1. Similarly, at high inclinations the smaller area due to oblateness and the additional redshift due to the quadrupole moment outweigh the Doppler boost to cause a decrease in the flux. As expected from the order-of-magnitude estimates above, the dominant effect is that of stellar oblateness.

 \begin{figure}
\psfig{file=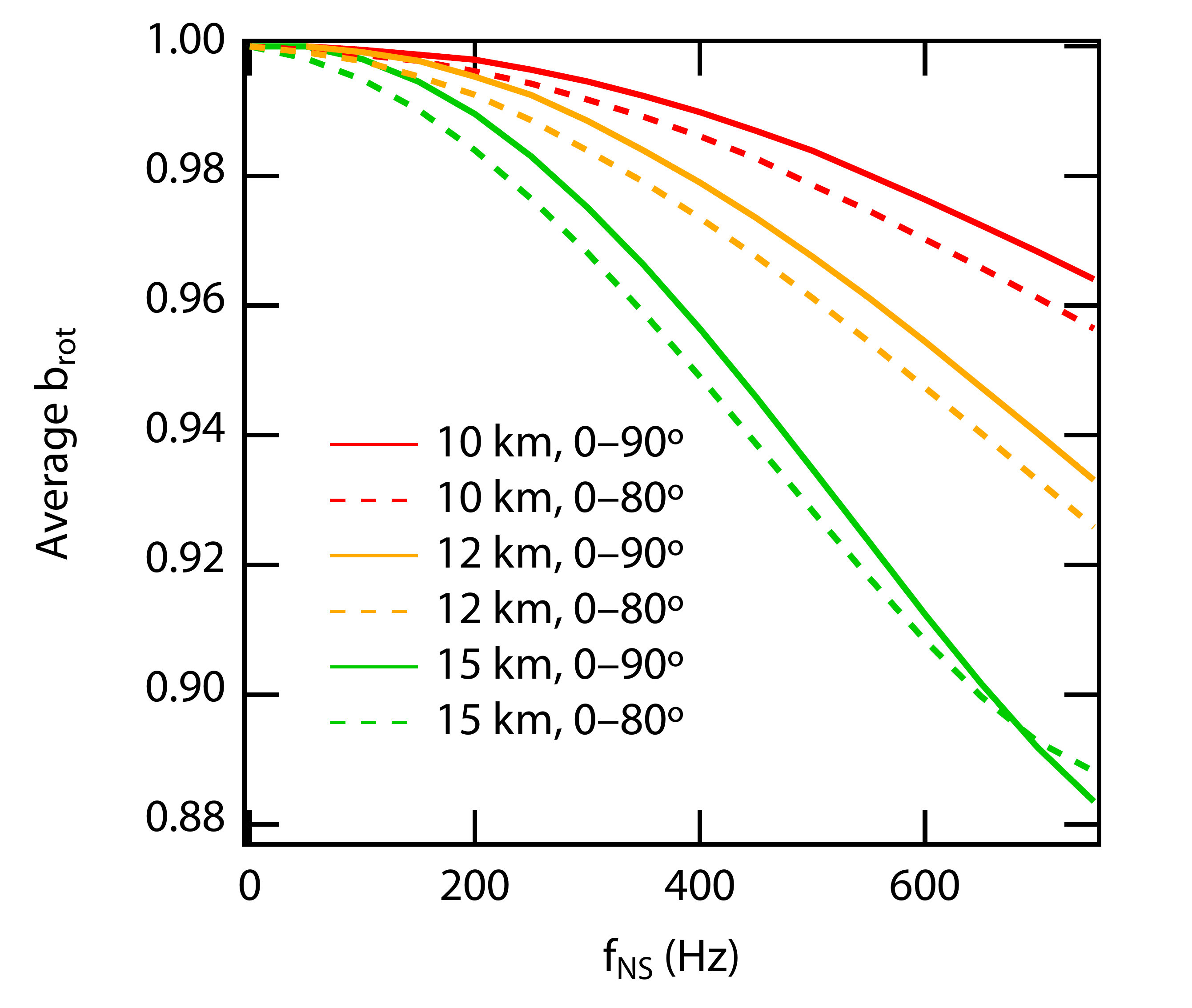, width=3.5in}
\caption{Rotational correction to the bolometric flux averaged over inclination. As in Figure~\ref{fig:fc_sinth}, the three colored lines show the averaged correction for neutron stars with three different radii. Solid lines correspond to the correction averaged over 0 to 90 degrees, while dashed lines show the correction averaged over 0 to 80 degrees. In the case of a neutron star with a radius of 12~km, the solid and dashed lines are indistinguishable.}
\label{fig:N_sinth}
\end{figure}   
 
In Figure~\ref{fig:N_sinth}, we show the correction to the bolometric flux averaged over the inclination angle. As in Figure~\ref{fig:fc_sinth}, we use two inclination ranges to account for possible obscuration of the source by an accretion disk. We find that the flux observed from a neutron star spinning above 600~Hz is 2\% to 10\% lower than for a non-spinning source. The largest flux corrections apply to neutron stars with the larger radii, which are more strongly modified by the second-order effects. In general, excluding the highest inclination sources from the average leads to a smaller value of $b_{\rm rot}$ (i.e., a value that is further from 1). This is because the smaller inclination range excludes the high-inclination sources for which the increased gravitational redshift due to the oblate shape of the surface plays only a small role. Without these sources, the average flux correction factor decreases.  For very large radii and high spin frequencies, the significantly oblate surface decreases the emitting area of the star and consequently the flux correction. If these high-inclination sources are obscured, the inclination-averaged flux correction factor increases.

\begin{figure}
\psfig{file=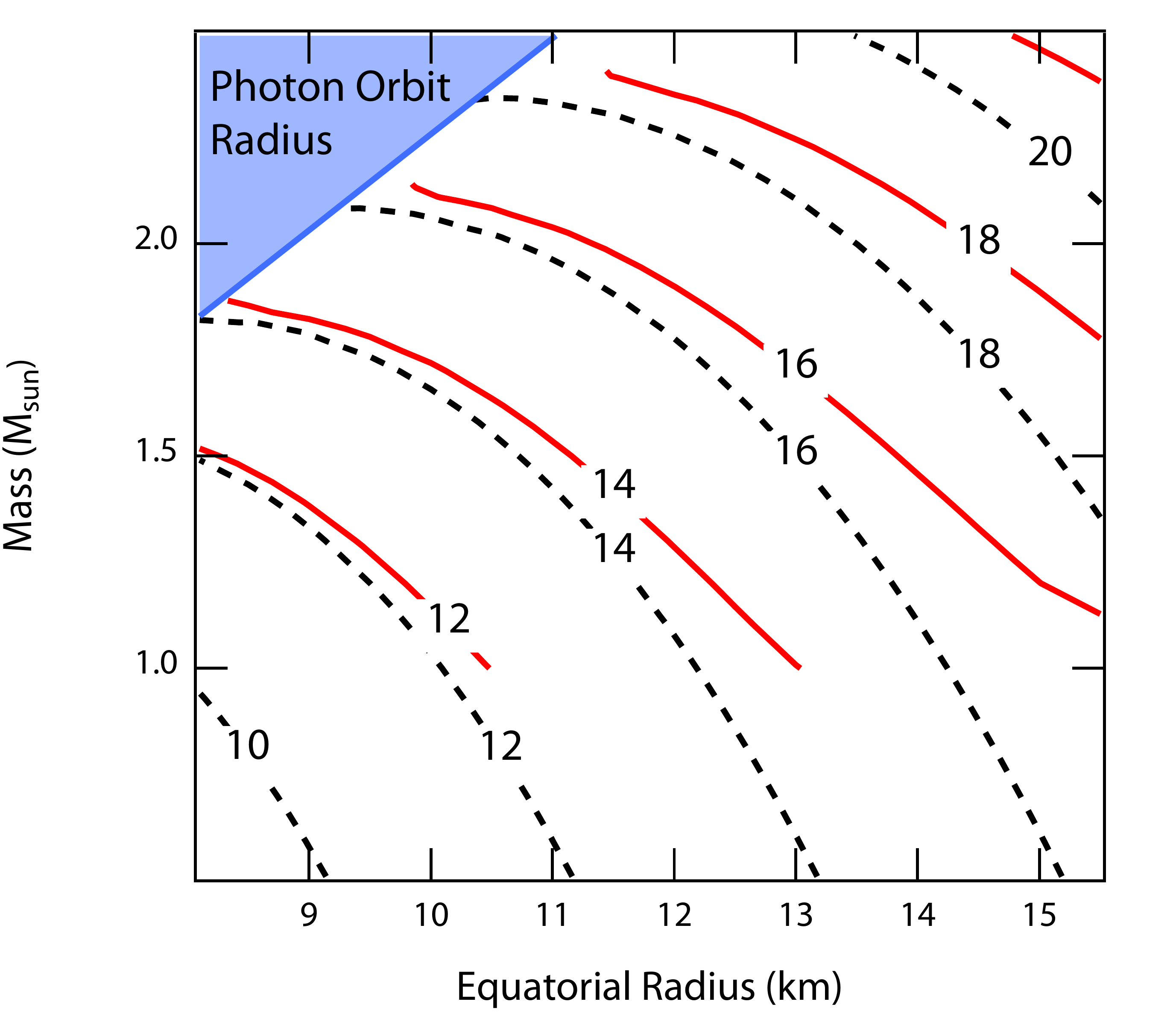, width=3.5in}
\caption{Contours of constant spectroscopic radius $R_\infty$ plotted over the parameter space of equatorial radius and mass. The dashed lines are the contours for a non-spinning Schwarzschild star defined by equation~(\ref{eq:sch_r}). The solid lines are the contours defined by equation~(\ref{eq:R_correction}) for stars spinning at 600~Hz averaged over all inclination angles. The shaded region in the upper left corner corresponds to masses and radii for which the neutron star surface is within the photon orbit. Over the whole parameter space, a spinning star shows a smaller apparent radius than a non-spinning star for the same value of mass and equatorial radius. }
\label{fig:M_R_App}
\end{figure}

$\\$

\section{Conclusions}

We have shown that moderate spins have significant effects on the observed thermal spectra from neutron-star surfaces. At spin frequencies above 300~Hz, terms of second order in $v/c$ can change observable quantities such as the flux and surface temperature. For moderately spinning stars, the temperature and flux measured by an observer at infinity is biased by factors related to the spin and the gravitational redshift of the stellar surface. Whether the inferred temperature is higher or lower than that of a non-spinning source depends on the parameters of the neutron star. However, we find that the magnitude of the bias is less than 2\% across the parameter space. Increasing the spin of the source also results in a decrease in the flux observed at infinity. The magnitude of this decrease can be up to 12\% for realistic neutron-star parameters. In general, neutron stars with larger radii experience stronger rotational effects and therefore have a decreased flux.

Given these biases, we can calculate how the correction to the flux and temperature due to the rapid spin of the source translates into a correction on the radius measurement. In Figure~\ref{fig:M_R_App}, we show the apparent spectroscopic radius measured for a non-spinning star and one spinning at 600~Hz for a range of masses and radii. The dashed lines show the apparent radii for non-spinning stars in the Schwarzschild metric given in equation~(\ref{eq:sch_r}). The solid lines correspond to sources spinning at 600~Hz and is given by equation~(\ref{eq:R_correction}). The reduction in the flux that reaches the observer at infinity leads to a smaller apparent radius across the parameter space. In particular, assuming a non-spinning source leads to significant underestimation of the equatorial radius in the case of neutron stars with large radii and small masses.

Consider a star that spins at a frequency of 600~Hz and is known to have a mass of 1.4~M$_\odot$. If a measurement yields an apparent spectroscopic radius of 14~km and one uses the Schwarzschild approximation to calculate the equatorial radius, the dashed curve labeled 14 gives a value of 10.8~km. Using the solid curve which corrects for the effects of the spin yields an equatorial radius of 11.2~km. In this case, using the non-spinning approximation underestimates the radius by 3.5\%. 

\begin{figure}
\psfig{file=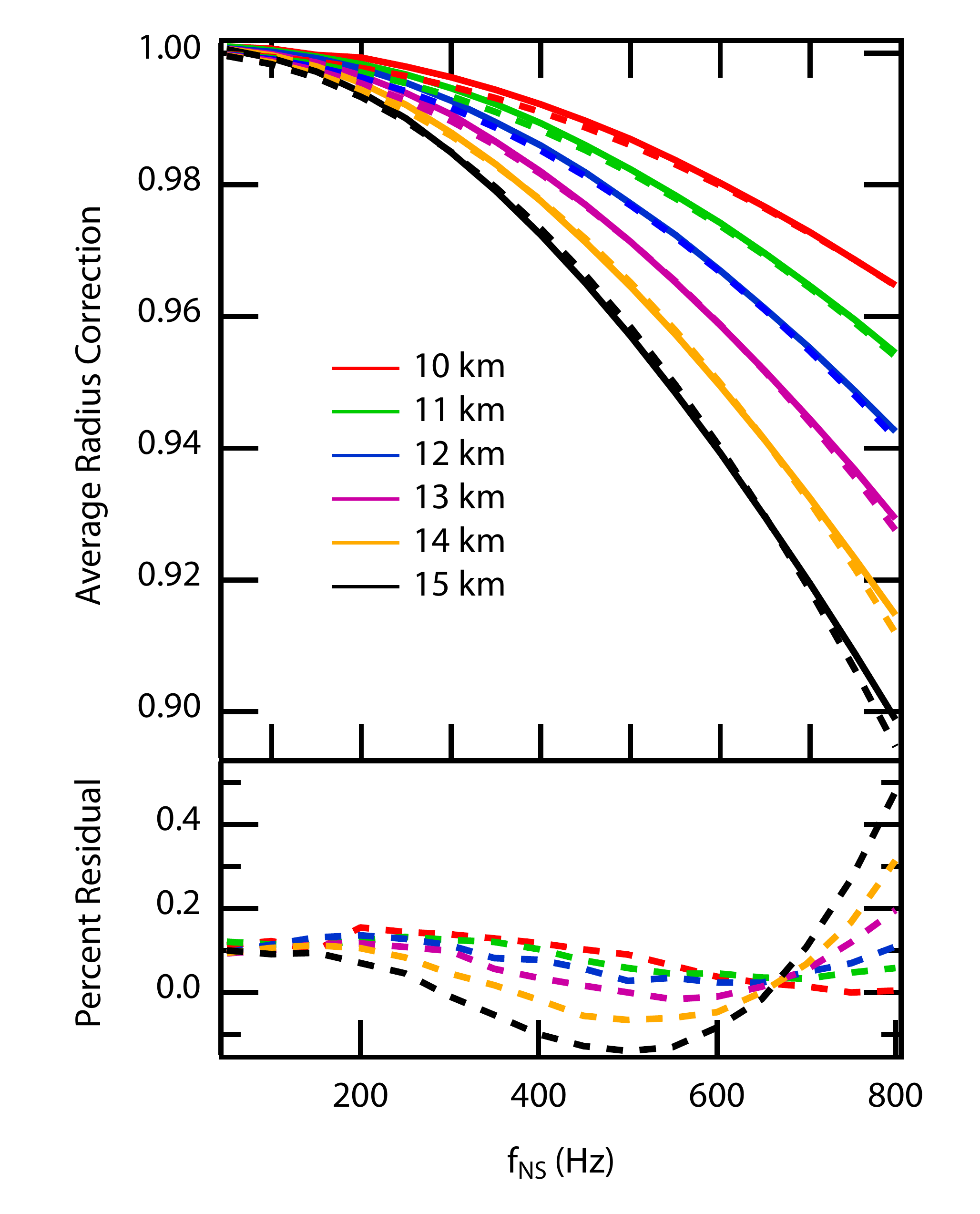, width=3.5in}
\caption{Rotational correction to the spectroscopically measured radii averaged over the inclination of the observer. The solid lines show the correction for different neutron-star radii, as in Figure~\ref{fig:N_sinth}. Dashed lines show empirical fits corresponding to equation~(\ref{eq:R_app_fit}). The lower panel shows the percent residual in each fit.}
\label{fig:b_rot_fits}
\end{figure}

Finally, we find an empirical formula to correct the bias introduced to radius measurements by the assumption that a source is non-spinning. We calculate the bolometric flux correction factor $b_{\rm rot}$ and temperature correction factor $\zeta_{\rm rot}$ for neutron stars with masses between 1.1 and 2.0~$M_\odot$ and radii between 10 and 15~km for all inclination angles and spin frequencies up to 800~Hz. As in Figures \ref{fig:fc_sinth} and \ref{fig:N_sinth}, we average these correction factors over inclination. Using equation~(\ref{eq:R_correction}), we can then find the bias in the apparent radius inferred from the measured flux and temperature. The solid lines in the upper panel of Figure~\ref{fig:b_rot_fits} show these correction factors. We calculate least-squares quadratic fits to the correction factor as a function of spin frequency for each neutron-star radius. Lastly, we fit the coefficients of this quadratic as a function of the radius to find a general equation for the spectroscopic radius correction,  
\begin{multline}
\frac{R_\infty^{f_{\rm NS}}}{R_{\infty}^{f_{\rm NS} = 0}} = 1 + \bigg[ \left(0.108 - 0.096 \frac{M}{M_\odot} \right) + \\
\left(-0.061 + 0.114 \frac{M}{M_\odot} \right) \frac{R}{10~\mathrm{km}}  -\\
0.128 \left(\frac{R}{10~\mathrm{km}}\right)^2 \bigg]  \left(\frac{f_{\rm NS}}{1000 \, \mathrm{Hz}}\right)^2.
\label{eq:R_app_fit}
\end{multline}
The dashed lines in Figure~\ref{fig:b_rot_fits} show this fit for 1.4~M$_\odot$ neutron stars with six chosen values of radius. The lower panel of Figure~\ref{fig:b_rot_fits} shows the residuals to this fit. For neutron stars with radii between 10 and 15~km and masses between 1.1 and 2.0~M$_\odot$, we find that this fit approximates the numerical value of $R_\infty^{f_{\rm NS}}/R_\infty^{f_{\rm NS} = 0}$ to within 0.5\% for spin frequencies below 800~Hz. 

Our results have two important implications for the spectroscopic measurements of neutron-star masses and radii in X-ray bursters and quiescent sources. First, the distortion of the observed spectra from blackbodies are predicted to be at the $\simeq 5$\% level. This is comparable to the spectral distortions caused by atmospheric effects (see, e.g. Suleimanov et al.\ 2011) and to the inferred deviation of the observed spectra in thermonuclear bursters from blackbodies (G\"uver et al.\ 2012). Second, the equatorial radii of neutron stars spinning at moderate rates inferred under the assumption that the stars are non-spinning are underestimated at a similar level. The bias depends on the spin and radius of the neutron star as well as on the observer's inclination but can be corrected with the formulae presented in this work. 

\acknowledgements
This research at the University of Arizona was supported by NSF grants AST~1108753 and AST~1312034, Chandra Theory grant TM2-13002X, and a grant from NSERC to S.M.M. F.\"O. thanks the Miller Institute for Basic Research in Science, University of California Berkeley for their support and hospitality. In addition, S.M.M. wishes to thank Steward Observatory for hospitality during her sabbatical visit.

\end{document}